# Comment on Jackson's analysis of electric charge quantization due to interaction with Dirac's magnetic monopole

Masud Mansuripur

College of Optical Sciences, The University of Arizona, Tucson, Arizona 85721

**Abstract**. In J. D. Jackson's *Classical Electrodynamics* textbook, the analysis of Dirac's charge quantization condition in the presence of a magnetic monopole has a mathematical omission and an all too brief physical argument that might mislead some students. This paper presents a detailed derivation of Jackson's main result, explains the significance of the missing term, and highlights the close connection between Jackson's findings and Dirac's original argument.

**1. Introduction**. In a 1931 paper [1], Dirac showed that the existence of a single magnetic monopole in the universe suffices to explain the observed discreteness of electrical charge. Dirac postulated a magnetic monopole residing at the terminus of a semi-infinite, uniformly-magnetized string, as shown in Fig. 1. He thus ensured the satisfaction of Maxwell's equation $\nabla \cdot \boldsymbol{B} = 0$, which is a prerequisite for defining the vector potential $\boldsymbol{A}$ via the identity $\nabla \times \boldsymbol{A} = \boldsymbol{B}$.

Dirac's subtle argument relies on the strong vorticity of the vector potential surrounding the string, which prevents the quantum-mechanical wave-function of an electrically-charged particle (e.g., an electron) from penetrating the string; the string thus remains invisible to the electrically-charged particle. However, for the phase of the wave-function in the vicinity of the string to be single-valued, Dirac argued that the product of the particle's electric charge $q$ and the monopole's magnetic charge $m_0$ must be an integer-multiple of Planck's constant $h$. (**Note**: Dirac's quantization condition is $q m_0 = nh$ in the *SI* system of units, and $4\pi q m_0 = nhc$ in the Gaussian system; here $h$ is Planck's constant, $c$ is the speed of light in vacuum, and $n$ is an arbitrary nonzero integer.)

The presentation of Dirac's argument in J. D. Jackson's *Classical Electrodynamics* textbook [2] deviates somewhat from Dirac's original line of reasoning. Jackson arrives at the correct charge quantization condition despite a mathematical omission and a rather hasty physical argument. There is much to recommend Jackson's analysis of Dirac's quantization condition, and it would be a pity if a minor omission and a hasty shortcut distracted the reader from fully appreciating the significance of this analysis. The present paper aims to expand upon and clarify Jackson's discussion of charge quantization in the presence of a Dirac magnetic monopole.

Following the publication of Dirac's famous 1931 paper, there have appeared many books and papers that elaborate and expand upon Dirac's ideas. There have also been several attempts at capturing and detecting magnetic monopoles. The short list of cited references here [3-7] is by no means intended to provide a comprehensive guide to the vast literature of the subject. We do hope, however, that, upon consulting these references, the interested reader will catch a glimpse of where the studies of magnetic monopoles stand today.

**2. String's vector potential**. With reference to Fig. 1, a semi-infinite magnetized string (or solenoid) may be modelled as follows:

$$\boldsymbol{M}(\boldsymbol{r}) = \boldsymbol{m}(z)\delta[x - x_s(z)]\delta[y - y_s(z)]. \qquad (1)$$

Here $m(z)$ is the magnetic dipole moment *per unit length* of the string at $[x_s(z), y_s(z), z]$, and $M(r)$ is the magnetization density at $r = (x, y, z)$. There are no magnetic charges anywhere along the length of the string except at its extremities. As shown in Appendix A, the z-component of $m(z)$ must be a constant, that is, $m_z(z) = m_0$, where $m_0$ is the charge of the magnetic monopole at the terminal point $[x_s(z_0), y_s(z_0), z_0]$ of the string. (Here the magnetic induction $B$, the magnetic field $H$, and the magnetization $M$ are related via $B = \mu_0 H + M$, where $\mu_0$ is the permeability of free space. Both $B$ and $M$ thus have units of weber/m$^2$, resulting in $m_0$ being in webers.)

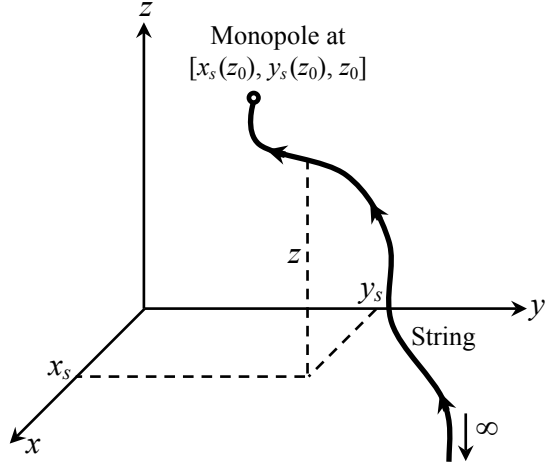

**Figure 1**. Semi-infinite magnetized string ending at a magnetic monopole at $z = z_0$. The coordinates of the string at elevation $z$ are given by the functions $x_s(z)$ and $y_s(z)$. The magnetic dipole moment *per unit length* of the string at $[x_s(z), y_s(z), z]$ is given by $m(z)$, which is aligned with the local orientation of the string and vanishes above $z = z_0$. The polar coordinates $\theta(z)$ and $\phi(z)$ of the string are related to its Cartesian coordinates via the identities $x'_s(z) = \tan\theta \cos\phi$ and $y'_s(z) = \tan\theta \sin\phi$. The alignment of $m(z)$ with the local orientation of the string thus implies that $m_x(z)/m_z(z) = x'_s(z)$ and $m_y(z)/m_z(z) = y'_s(z)$. The model used for the string in Eq.(1) gives the string a constant cross-section in the $xy$-plane. To ensure that magnetic charge does not appear anywhere along the string (except at $z = z_0$), the z-component of $m(z)$ must be a constant, that is, $m_z(z) = m_0$. The magnetic charge of the monopole at $[x_s(z_0), y_s(z_0), z_0]$ is thus equal to $m_0$, which, in the *SI* system, has units of weber.

To confirm that arbitrary bends and twists of a tightly-wound solenoidal string do not leak its internal magnetic flux to the outside world, a systematic calculation of the string's vector potential is presented in Appendix A. In the end, our derivation yields the same simple formula for the string's vector potential as that used by Jackson [2], namely, a direct integral over infinitesimal vector potentials at the observation point $r$ contributed by small lengths $d\ell$ of the string located at $r_s$, as follows:

$$A_S(r) = \frac{m_0}{4\pi} \int_\infty^{z_0} (\nabla_r |r - r_s|^{-1}) \times d\ell. \tag{2}$$

As a simple example (and one that was analyzed in detail by Dirac), we calculate $A_{S_1}$ for a monopole at the end of a string $S_1$ aligned with the negative z-axis. The monopole $m_0$ is thus located at the origin of coordinates, and the observation point $r = x\hat{x} + y\hat{y} + z\hat{z}$ has spherical coordinates $(r, \theta, \phi)$. A straightforward evaluation of Eq.(2) yields

$$\boldsymbol{A}_{S_1}(\boldsymbol{r}) = \tfrac{m_0}{4\pi} \int_{-\infty}^{0} \{\nabla_r [x^2 + y^2 + (z-\tilde{z})^2]^{-\frac{1}{2}}\} \times \hat{\boldsymbol{z}} \, d\tilde{z}$$

$$= -\tfrac{m_0}{4\pi} \int_{-\infty}^{0} \tfrac{[x\hat{x} + y\hat{y} + (z-\tilde{z})\hat{z}] \times \hat{z}}{[x^2+y^2+(z-\tilde{z})^2]^{3/2}} \, d\tilde{z} = \tfrac{m_0(x\hat{y}-y\hat{x})}{4\pi(x^2+y^2)} \int_{\cot\theta}^{\infty} \tfrac{d\zeta}{(1+\zeta^2)^{3/2}}$$

$$= \tfrac{m_0 \hat{\phi}}{4\pi\sqrt{x^2+y^2}} \times \tfrac{\zeta}{\sqrt{1+\zeta^2}}\Big|_{\cot\theta}^{\infty} = \tfrac{m_0 \hat{\phi}}{4\pi r \sin\theta}\left(1 - \tfrac{\cot\theta}{\sqrt{1+\cot^2\theta}}\right) = \tfrac{m_0(1-\cos\theta)\hat{\phi}}{4\pi r \sin\theta}. \qquad (3)$$

The term $(1 - \cos\theta)/\sin\theta$ may be further simplified and written as $\tan(\tfrac{1}{2}\theta)$. The vector potential of a semi-infinite string aligned with the negative z-axis is thus seen to have a singularity at the location of the string, i.e., at $\theta = \pi$. The vortex-like $\boldsymbol{A}_{S_1}$ circling the negative z-axis integrates to $m_0$ over a tight circle surrounding the string, in accordance with $\oint \boldsymbol{A}_{S_1} \cdot d\boldsymbol{\ell} = \int \boldsymbol{B}_{S_1} \cdot d\boldsymbol{\sigma} = m_0$. Now, Dirac's original argument in [1] appears to have been that the integral of $\boldsymbol{A}_{S_1}$ around a tight loop circling the string, when multiplied by $e/\hbar$, is a phase-factor for the Schrödinger wave-function $\psi_{S_1}(\boldsymbol{r})$ of an electron of charge $e$ in the presence of the magnetic monopole, which must be equal to an integer-multiple of $2\pi$. Dirac's quantization condition, $em_0/\hbar = 2\pi n$, is an immediate consequence of this assumption. (Here, as usual, $\hbar = h/2\pi$.)

Suppose now that a string $S_2$ extends from $z = 0$ to $+\infty$ along the positive z-axis, with the magnetic monopole $m_0$ residing at its lower terminus. A similar analysis as in Eq.(3) now yields

$$\boldsymbol{A}_{S_2}(\boldsymbol{r}) = \tfrac{m_0(-1-\cos\theta)\hat{\phi}}{4\pi r \sin\theta}. \qquad (4)$$

This time, the vector potential has vortex-like behavior around the positive z-axis, but, elsewhere in space, it produces the same $B$-field as does $\boldsymbol{A}_{S_1}$. The difference between the two vector potentials in Eqs.(3) and (4) is readily seen to be

$$\boldsymbol{A}_{S_1}(\boldsymbol{r}) - \boldsymbol{A}_{S_2}(\boldsymbol{r}) = \tfrac{m_0 \hat{\phi}}{2\pi r \sin\theta} = \nabla\left(\tfrac{m_0}{2\pi}\phi\right). \qquad (5)$$

The two vector potentials thus differ by the gradient of a scalar function, which might indicate that they are related via a gauge transformation. Note, however, that along the entire z-axis, the scalar function $m_0\phi/2\pi$ is ill-defined. Consequently, the z-axis is a singularity of the gradient of the scalar function. The difference between $\boldsymbol{A}_{S_1}$ and $\boldsymbol{A}_{S_2}$ is more than a *simple* gauge transformation; the curl of $\boldsymbol{A}_{S_1} - \boldsymbol{A}_{S_2}$ is the infinite magnetization inside a long, thin string extending all the way from $z = -\infty$ to $z = +\infty$.

**3. Change in vector potential in consequence of a change of the string**. In Jackson's treatment [3] of Dirac's magnetic monopole, we are reminded that the specific shape and/or location of the string is irrelevant and that, therefore, the vector potentials $\boldsymbol{A}_{S_1}$ and $\boldsymbol{A}_{S_2}$ corresponding to two strings $S_1$ and $S_2$, which terminate on the same monopole, must differ by a gauge transformation; see Fig.2. This means that $\boldsymbol{A}_{S_1}(\boldsymbol{r}) - \boldsymbol{A}_{S_2}(\boldsymbol{r}) = \nabla\chi(\boldsymbol{r})$, where $\chi(\boldsymbol{r})$ is some well-defined function of the spatial coordinates. Jackson proceeds to determine $\chi(\boldsymbol{r})$ along the following lines, but, toward the end, he appears to have inadvertently omitted a term containing a $\delta$-function. (Justifications for some of the steps taken below are given in Appendix B.)

$$\boldsymbol{A}_{S_1}(\boldsymbol{r}) - \boldsymbol{A}_{S_2}(\boldsymbol{r}) = \tfrac{m_0}{4\pi} \oint_C (\nabla_r |\boldsymbol{r} - \boldsymbol{r}_s|^{-1}) \times d\boldsymbol{\ell} \qquad (6a)$$

$$= \frac{m_0}{4\pi} \oint_c \nabla_r \times \frac{d\boldsymbol{\ell}}{|\boldsymbol{r}-\boldsymbol{r}_s|} \tag{6b}$$

$$= \frac{m_0}{4\pi} \nabla_r \times \oint_c \frac{d\boldsymbol{\ell}}{|\boldsymbol{r}-\boldsymbol{r}_s|} \tag{6c}$$

$$= -\frac{m_0}{4\pi} \nabla_r \times \int_s (\nabla_{r_s}|\boldsymbol{r}-\boldsymbol{r}_s|^{-1}) \times d\boldsymbol{\sigma} \tag{6d}$$

$$= \frac{m_0}{4\pi} \nabla_r \times \int_s (\nabla_r|\boldsymbol{r}-\boldsymbol{r}_s|^{-1}) \times d\boldsymbol{\sigma} \tag{6e}$$

$$= \frac{m_0}{4\pi} \nabla_r \times \nabla_r \times \int_s \frac{d\boldsymbol{\sigma}}{|\boldsymbol{r}-\boldsymbol{r}_s|} \tag{6f}$$

$$= \frac{m_0}{4\pi} \nabla_r \int_s \nabla_r \cdot \left(\frac{d\boldsymbol{\sigma}}{|\boldsymbol{r}-\boldsymbol{r}_s|}\right) - \frac{m_0}{4\pi} \int_s \nabla_r^2 \left(\frac{d\boldsymbol{\sigma}}{|\boldsymbol{r}-\boldsymbol{r}_s|}\right) \tag{6g}$$

$$= \frac{m_0}{4\pi} \nabla_r \int_s (\nabla_r|\boldsymbol{r}-\boldsymbol{r}_s|^{-1}) \cdot d\boldsymbol{\sigma} - \frac{m_0}{4\pi} \int_s (\nabla_r^2|\boldsymbol{r}-\boldsymbol{r}_s|^{-1}) d\boldsymbol{\sigma} \tag{6h}$$

$$= \frac{m_0}{4\pi} \nabla_r \int_s \frac{(\boldsymbol{r}_s - \boldsymbol{r}) \cdot d\boldsymbol{\sigma}}{|\boldsymbol{r}-\boldsymbol{r}_s|^3} + m_0 \int_s \delta_3(\boldsymbol{r}-\boldsymbol{r}_s) d\boldsymbol{\sigma} \tag{6i}$$

$$= \frac{m_0}{4\pi} \nabla_r \int_s d\Omega(\boldsymbol{r};\boldsymbol{r}_s) + m_0 \int_s \delta_3(\boldsymbol{r}-\boldsymbol{r}_s) d\boldsymbol{\sigma} \tag{6j}$$

$$= \frac{m_0}{4\pi} \nabla\Omega_c(\boldsymbol{r}) + m_0 \delta(r_\perp) \hat{\boldsymbol{r}}_\perp. \tag{6k}$$

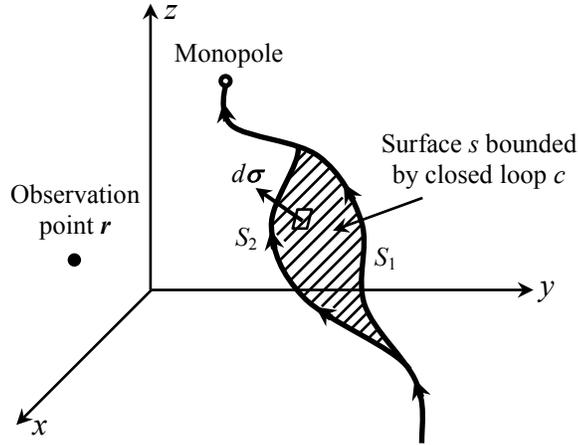

**Figure 2**. When two semi-infinite strings $S_1$ and $S_2$ terminate on the same magnetic monopole, the difference $\boldsymbol{A}_{S_1} - \boldsymbol{A}_{S_2}$ between their vector potentials is given by an integral over the closed loop $c$, which is the boundary of the shaded area $s$. The loop in this figure is traversed counterclockwise. The solid angle subtended by the closed loop $c$ at the observation point $\boldsymbol{r}$ is denoted by $\Omega_c(\boldsymbol{r})$. The points on the surface $s$ (and its boundary $c$) are denoted by $\boldsymbol{r}_s$. An elemental surface area of $s$ is denoted by $d\boldsymbol{\sigma}$, whose direction, while perpendicular to the local surface, is also tied to the direction of travel around $c$ via the right-hand rule. The elemental solid angle subtended at the observation point $\boldsymbol{r}$ by the surface element $d\boldsymbol{\sigma}$ at $\boldsymbol{r}_s$ is denoted by $d\Omega(\boldsymbol{r};\boldsymbol{r}_s)$.

In the preceding equation, $\Omega_c(\boldsymbol{r})$ is the solid angle subtended by the contour $c$ when viewed from $\boldsymbol{r}$, and $r_\perp$ is the perpendicular distance from $\boldsymbol{r}$ to the surface $s$. Since $\delta(r_\perp)$ is nonzero only when $\boldsymbol{r}$ is extremely close to $s$, the unit-vector $\hat{\boldsymbol{r}}_\perp$ coincides with the local surface normal, its orientation being determined by the sense of travel around $c$. The sign of $r_\perp$ is positive or negative, depending on which side of the surface $s$ (as indicated by the direction of $\hat{r}_\perp$) the observation point $r$ happens to fall.

In Jackson's analysis [2], the term $m_0 \delta(r_\perp)\hat{r}_\perp$ on the right-hand-side of Eq.(6k) is missing. This is the aforementioned mathematical omission, whose effects are subsequently compounded by the brevity of Jackson's physical argument. Jackson notes that, when the observation point $r$ crosses the surface $s$ along the surface-normal $\hat{r}_\perp$, the solid angle $\Omega_c(r)$ suddenly drops from $2\pi$ to $-2\pi$. This discontinuity of $\Omega_c(r)$ at the surface $s$ gives rise to a $\delta$-function of magnitude $4\pi$ whenever $\nabla\Omega_c(r)$ is evaluated at a point $r$ located on the surface $s$. However, the resulting $\delta$-function contained in $(m_0/4\pi)\nabla\Omega_c(r)$ is readily cancelled out by the term $m_0\delta(r_\perp)\hat{r}_\perp$ appearing in Eq.(6k). This means that $A_{S_1} - A_{S_2}$ is a well-defined, continuous function of $r$ everywhere except on the closed loop $c$ (i.e., on the boundary of $s$). It is worth emphasizing that, while $\Omega_c(r)$ is inherently discontinuous at the surface $s$, the vector-potentials $A_{S_1}(r)$ and $A_{S_2}(r)$ are expected to be smooth and well-behaved functions of $r$ everywhere in space — except, of course, on their respective strings $S_1$ and $S_2$. Consequently, the presence of $m_0\delta(r_\perp)\hat{r}_\perp$ in Eq.(6k) is absolutely essential if the $\delta$-function contained in $\nabla\Omega_c(r)$ is to be neutralized.

At this point, one is not yet in a position to specify a gauge, because $\Omega_c(r)$, a function that is well-defined everywhere except on $c$, has a $4\pi$ discontinuity on $s$. This discontinuity gives rise to a $\delta$-function in $\nabla\Omega_c$, which is removed only after an equal and opposite $\delta$-function is added to $\nabla\Omega_c$. One must somehow eliminate, or render invisible, the discontinuity of $\Omega_c$ at $s$, which is responsible for the *undesirable* $\delta$-function appearing in Eq.(6k). Removal of this discontinuity requires that $\Omega_c$ be incorporated into a phase-factor, as explained in the next section.

**4. Schrödinger's equation for point-charge in the presence of magnetic monopole**. Suppose $\psi_{S_2}(r,t)$ is a solution of Schrödinger's equation for a point-particle of charge $q$ and mass $m$ in the presence of the string $S_2$, whose vector potential is $A_{S_2}$. Let us multiply the wave-function $\psi_{S_2}$ by the *spurious* phase-factor $\exp[iqm_0\Omega_c(r)/(4\pi\hbar)]$. This phase-factor will be discontinuous at the surface $s$, where $\Omega_c$ has a $4\pi$ jump, unless $qm_0/\hbar$ happens to be an integer-multiple of $2\pi$. Therefore, imposing Dirac's quantization condition $qm_0 = 2\pi n\hbar = nh$ renders *invisible* the discontinuity of $\Omega_c$ at $s$. Under such circumstances, when evaluating the gradient of the phase-factor, one is obligated to add the necessary $\delta$-function to $\nabla\Omega_c$ in order to ensure that the resulting function is well-behaved, that is,

$$-i\hbar\nabla\exp\left(\frac{iqm_0\Omega_c}{4\pi\hbar}\right) = \frac{qm_0}{4\pi}[\nabla\Omega_c + 4\pi\delta(r_\perp)\hat{r}_\perp]\exp\left(\frac{iqm_0\Omega_c}{4\pi\hbar}\right). \tag{7}$$

Substitution from Eq.(6) into the above equation now yields

$$-i\hbar\nabla\exp\left(\frac{iqm_0\Omega_c}{4\pi\hbar}\right) = q(A_{S_1} - A_{S_2})\exp\left(\frac{iqm_0\Omega_c}{4\pi\hbar}\right). \tag{8}$$

Note how the imposition of Dirac's quantization condition has "patched up" the discontinuity of the phase-factor in such a way as to render $\Omega_c$ *effectively* continuous at the surface $s$. Had the $\delta$-function in Eq.(6k) been absent, there would have been no need in Jackson's analysis [2] for Dirac's quantization condition, because the function $(m_0/4\pi)\Omega_c(r)$, in spite of its discontinuity at $s$, would have been an acceptable gauge.

It is now easy to show, with the aid of Eq.(8), that the product of $\psi_{S_2}$ and the *spurious* phase-factor satisfies Schrödinger's equation for the point-particle of charge $q$ and mass $m$ in the presence of $\boldsymbol{A}_{S_1}$, the vector potential of $S_1$. [To simplify the notation, we pretend in what follows that $\Omega_c$ is continuous at the surface $s$, and proceed to omit the $\delta$-function that, in accordance with Eq.(7), must accompany $\boldsymbol{\nabla}\Omega_c$.] We write

$$\begin{aligned}
\frac{1}{2m}(-i\hbar\boldsymbol{\nabla} - q\boldsymbol{A}_{S_1}) \cdot (-i\hbar\boldsymbol{\nabla} - q\boldsymbol{A}_{S_1})&\left[\exp\left(\frac{iqm_0\Omega_c}{4\pi\hbar}\right)\psi_{S_2}\right]\\
&= \frac{1}{2m}(-i\hbar\boldsymbol{\nabla} - q\boldsymbol{A}_{S_1}) \cdot \left[-i\hbar\boldsymbol{\nabla}\psi_{S_2} - q\left(\boldsymbol{A}_{S_1} - \frac{m_0}{4\pi}\boldsymbol{\nabla}\Omega_c\right)\psi_{S_2}\right]\exp\left(\frac{iqm_0\Omega_c}{4\pi\hbar}\right)\\
&= \frac{1}{2m}(-i\hbar\boldsymbol{\nabla} - q\boldsymbol{A}_{S_1}) \cdot \left[(-i\hbar\boldsymbol{\nabla}\psi_{S_2} - q\boldsymbol{A}_{S_2}\psi_{S_2})\exp\left(\frac{iqm_0\Omega_c}{4\pi\hbar}\right)\right]\\
&= \frac{1}{2m}\left\{\left[-i\hbar\boldsymbol{\nabla} - q\left(\boldsymbol{A}_{S_1} - \frac{m_0}{4\pi}\boldsymbol{\nabla}\Omega_c\right)\right] \cdot (-i\hbar\boldsymbol{\nabla}\psi_{S_2} - q\boldsymbol{A}_{S_2}\psi_{S_2})\right\}\exp\left(\frac{iqm_0\Omega_c}{4\pi\hbar}\right)\\
&= \frac{1}{2m}[(-i\hbar\boldsymbol{\nabla} - q\boldsymbol{A}_{S_2}) \cdot (-i\hbar\boldsymbol{\nabla} - q\boldsymbol{A}_{S_2})\psi_{S_2}]\exp\left(\frac{iqm_0\Omega_c}{4\pi\hbar}\right)\\
&= (i\hbar\partial_t\psi_{S_2})\exp\left(\frac{iqm_0\Omega_c}{4\pi\hbar}\right)\\
&= i\hbar\partial_t\left[\exp\left(\frac{iqm_0\Omega_c}{4\pi\hbar}\right)\psi_{S_2}\right]. \quad (9)
\end{aligned}$$

Clearly, $\exp[iqm_0\Omega_c/(4\pi\hbar)]\,\psi_{S_2}$ is a solution of Schrödinger's equation for the charge $q$ in the presence of $S_1$. However, multiplication by a phase-factor is physically meaningless, since it is equivalent to a change of gauge. The fact that the phase-factor is ill-defined over the contour $c$ does not seem to have any physical significance either, as the wave-functions always vanish on the corresponding strings. It should now be clear that the vector potential associated with one string, say $S_1$, can produce, aside from a spurious phase-factor, the solution to Schrödinger's equation for any other string, such as $S_2$, as well.

In his original paper [1], Dirac gives an explicit example in which the vector potential associated with a string along the negative $z$-axis produces two eigen solutions to Schrödinger's equation, namely, $\psi_{1a} = f(r)\cos(\theta/2)$ and $\psi_{1b} = f(r)\sin(\theta/2)\exp(i\phi)$. The first solution corresponds to the string whose presence along the negative $z$-axis has been assumed, while the second solution, aside from the spurious phase-factor $\exp(i\phi)$, represents an eigen wave-function associated with a string along the positive $z$-axis. Note that the function $\exp(i\phi)$ is ill-defined along the entire $z$-axis, which, in the present example, represents the contour $c$. [Note: There is a minus sign missing in Dirac's paper; the spurious phase-factor for $\psi_{1b}$ should in fact be $\exp(-i\phi)$, as can be readily checked by substitution into his equation (13).]

1. Dirac, P. A. M. "Quantised Singularities in the Electromagnetic Field," *Proc. Roy. Soc. London* **A133**(821), pp. 60-72 (1931).
2. Jackson, J. D., *Classical Electrodynamics*, 3rd edition, Wiley, New York (1999); see pages 278-280 and also problems 6.18 and 6.19.
3. Milton, K. A. "Theoretical and experimental status of magnetic monopoles," *Rep. Prog. Phys.* **69**(6), pp. 1637-1711 (2006).
4. Chuang, I., Durrer, R., Turok, N. and Yurke, B. "Cosmology in the laboratory: defect dynamics in liquid crystals," *Science* **251**(4999), pp. 1336-1342 (1991).
5. Fang, Z., Nagaosa, N., Takahashi, K.S., Asamitsu, A., Mathieu, R., Ogasawara, T., Yamada, H., Kawasaki, M., Tokura, Y. and Terakura, K. "The anomalous Hall effect and magnetic monopoles in momentum space," *Science* **302**(5642), pp. 92-95 (2003).
6. Castelnovo, C., Moessner, R. and Sondhi, S.L. "Magnetic monopoles in spin ice," *Nature* **451**, pp. 42-45 (2008).
7. Ray, M.W., Ruokokoski, E., Kandel, S., Möttönen, M. and D.S. Hall, "Observation of Dirac monopoles in a synthetic magnetic field," *Nature* **505**, pp. 657-660 (2014).

# Appendix A

We derive a formula for the vector potential $A_s(r)$ of the arbitrary string $S$ depicted in Fig.1, whose magnetization $M(r)$ is given by Eq.(1). It must be pointed out that, in our notation, $B = \mu_0 H + M$, and that, therefore, magnetic induction $B$ and magnetization $M$ have the same units (tesla or weber/m$^2$). Here $H$ is the magnetic field (ampere/m), and $\mu_0$ is the permeability of free space (henry/m). The fundamental constraint on $M(r)$ is that its magnetic charge-density along the length of $S$ must vanish, that is,

$$\nabla \cdot M(r) = m_x(z)\delta'[x - x_s(z)]\delta[y - y_s(z)] + m_y(z)\delta[x - x_s(z)]\delta'[y - y_s(z)]$$
$$+ m_z'(z)\delta[x - x_s(z)]\delta[y - y_s(z)] - m_z(z)x_s'(z)\delta'[x - x_s(z)]\delta[y - y_s(z)]$$
$$- m_z(z)y_s'(z)\delta[x - x_s(z)]\delta'[y - y_s(z)] = 0. \tag{A1}$$

Consequently, $x_s'(z) = m_x(z)/m_z(z)$, $y_s'(z) = m_y(z)/m_z(z)$, and $m_z'(z) = 0$, which yields $m_z(z) = m_0$. Note that the string's local magnetic moment $m(z)$, which may also be described in terms of the polar angles $[\theta(z), \phi(z)]$, is aligned with the string's local orientation $d\ell = (x_s' \hat{x} + y_s' \hat{y} + \hat{z})dz$. At the terminus of the string, where $z = z_0$, there is a sudden change in $m_z(z)$, from $m_0$ to zero. At this terminal point, therefore, $m_z'(z_0) = -m_0 \delta(z - z_0)$, yielding the magnetic charge-density of the string as follows:

$$\rho_m(r) = -\nabla \cdot M(r) = m_0 \delta[x - x_s(z_0)]\delta[y - y_s(z_0)]\delta(z - z_0). \tag{A2}$$

The strength of the magnetic monopole located at $[x_s(z_0), y_s(z_0), z_0]$ is thus seen to be $m_0$, which has the dimensions of magnetic-dipole-moment-per-unit-length. In the $SI$ system of units, $m_0$ is in webers. Note that while $m(z)$ must terminate at $z = z_0$, the functions $x_s(z)$ and $y_s(z)$ may continue indefinitely beyond the terminal point.

To find the vector potential of the semi-infinite string, we first calculate its bound current-density, namely,

$$J_{\text{bound}}(r) = \mu_0^{-1} \nabla \times M(r)$$
$$= \mu_0^{-1}\{\partial_y[m_z \delta(x - x_s)\delta(y - y_s)] - \partial_z[m_y \delta(x - x_s)\delta(y - y_s)]\}\hat{x}$$
$$+ \mu_0^{-1}\{\partial_z[m_x \delta(x - x_s)\delta(y - y_s)] - \partial_x[m_z \delta(x - x_s)\delta(y - y_s)]\}\hat{y}$$
$$+ \mu_0^{-1}\{\partial_x[m_y \delta(x - x_s)\delta(y - y_s)] - \partial_y[m_x \delta(x - x_s)\delta(y - y_s)]\}\hat{z}$$
$$= \mu_0^{-1}[m_z \delta(x - x_s)\delta'(y - y_s) - \partial_z(m_y)\delta(x - x_s)\delta(y - y_s)$$
$$+ m_y x_s' \delta'(x - x_s)\delta(y - y_s) + m_y y_s' \delta(x - x_s)\delta'(y - y_s)]\hat{x}$$
$$+ \mu_0^{-1}[\partial_z(m_x)\delta(x - x_s)\delta(y - y_s) - m_x x_s' \delta'(x - x_s)\delta(y - y_s)$$
$$- m_x y_s' \delta(x - x_s)\delta'(y - y_s) - m_z \delta'(x - x_s)\delta(y - y_s)]\hat{y}$$
$$+ \mu_0^{-1}[m_y \delta'(x - x_s)\delta(y - y_s) - m_x \delta(x - x_s)\delta'(y - y_s)]\hat{z}. \tag{A3}$$

The vector potential at an arbitrary point $r$ is obtained by integrating over the volume of space (coordinates denoted by $\tilde{r}$) which contains the bound current-density given by Eq.(A3). The distance between the observation point $r$ and an arbitrary source point $\tilde{r}$ is written

$$|r - \tilde{r}| = \sqrt{(x - \tilde{x})^2 + (y - \tilde{y})^2 + (z - \tilde{z})^2}. \tag{A4}$$

In what follows $\partial_{\tilde{x}}|\boldsymbol{r}-\tilde{\boldsymbol{r}}|^{-1}$ is replaced with $-\partial_x|\boldsymbol{r}-\tilde{\boldsymbol{r}}|^{-1}$ and $\partial_{\tilde{y}}|\boldsymbol{r}-\tilde{\boldsymbol{r}}|^{-1}$ with $-\partial_y|\boldsymbol{r}-\tilde{\boldsymbol{r}}|^{-1}$. When $\tilde{\boldsymbol{r}}$ happens to reside on the string, its distance to the observation point $\boldsymbol{r}$ will be

$$|\boldsymbol{r}-\boldsymbol{r}_s| = \sqrt{[x-x_s(\tilde{z})]^2 + [y-y_s(\tilde{z})]^2 + (z-\tilde{z})^2}. \tag{A5}$$

In this case, we will have

$$\partial_{\tilde{z}}|\boldsymbol{r}-\boldsymbol{r}_s|^{-1} = -[x_s'(\tilde{z})\partial_x + y_s'(\tilde{z})\partial_y + \partial_z]|\boldsymbol{r}-\boldsymbol{r}_s|^{-1}. \tag{A6}$$

The standard sifting properties of the $\delta$-function, namely, $\int_{-\infty}^{\infty} f(x)\delta(x-\tilde{x})dx = f(\tilde{x})$ and $\int_{-\infty}^{\infty} f(x)\delta'(x-\tilde{x})dx = -f'(\tilde{x})$, will be used in the following derivations. The vector potential of the semi-infinite string may now be written as follows:

$$\boldsymbol{A}_S(\boldsymbol{r}) = \frac{\mu_0}{4\pi}\iiint \frac{J_{\text{bound}}(\tilde{\boldsymbol{r}})}{|\boldsymbol{r}-\tilde{\boldsymbol{r}}|}d\tilde{x}d\tilde{y}d\tilde{z}$$

$$= \frac{\hat{\boldsymbol{x}}}{4\pi}\int\left\{m_z\partial_y|\boldsymbol{r}-\boldsymbol{r}_s|^{-1} - \frac{\partial_{\tilde{z}}(m_y)}{|\boldsymbol{r}-\boldsymbol{r}_s|} + m_y x_s'\partial_x|\boldsymbol{r}-\boldsymbol{r}_s|^{-1} + m_y y_s'\partial_y|\boldsymbol{r}-\boldsymbol{r}_s|^{-1}\right\}d\tilde{z}$$

$$+ \frac{\hat{\boldsymbol{y}}}{4\pi}\int\left\{\frac{\partial_{\tilde{z}}(m_x)}{|\boldsymbol{r}-\boldsymbol{r}_s|} - m_x x_s'\partial_x|\boldsymbol{r}-\boldsymbol{r}_s|^{-1} - m_x y_s'\partial_y|\boldsymbol{r}-\boldsymbol{r}_s|^{-1} - m_z\partial_x|\boldsymbol{r}-\boldsymbol{r}_s|^{-1}\right\}d\tilde{z}$$

$$+ \frac{\hat{\boldsymbol{z}}}{4\pi}\int(m_y\partial_x - m_x\partial_y)|\boldsymbol{r}-\boldsymbol{r}_s|^{-1}d\tilde{z}. \tag{A7}$$

The integral $\int \partial_{\tilde{z}}(m_x)|\boldsymbol{r}-\boldsymbol{r}_s|^{-1}d\tilde{z}$ can be evaluated using the method of integration by parts, namely,

$$\int_{\infty}^{z_0^+} \frac{\partial_{\tilde{z}}(m_x)}{|\boldsymbol{r}-\boldsymbol{r}_s|}d\tilde{z} = \frac{m_x(\tilde{z})}{|\boldsymbol{r}-\boldsymbol{r}_s|}\bigg|_{\infty}^{z_0^+} + \int_{\infty}^{z_0^+} m_x(\tilde{z})(x_s'\partial_x + y_s'\partial_y + \partial_z)|\boldsymbol{r}-\boldsymbol{r}_s|^{-1}d\tilde{z}. \tag{A8}$$

Note that the first term on the right-hand-side of Eq.(A8) vanishes at both ends of the string. A similar procedure can be applied to $\int \partial_{\tilde{z}}(m_y)|\boldsymbol{r}-\boldsymbol{r}_s|^{-1}d\tilde{z}$. Substitution into Eq.(A7) and using the fact that $m_x = x_s'm_z$ and $m_y = y_s'm_z$ now yields

$$\boldsymbol{A}_S(\boldsymbol{r}) = \frac{\hat{\boldsymbol{x}}}{4\pi}\int m_z(\partial_y - y_s'\partial_z)|\boldsymbol{r}-\boldsymbol{r}_s|^{-1}d\tilde{z} + \frac{\hat{\boldsymbol{y}}}{4\pi}\int m_z(x_s'\partial_z - \partial_x)|\boldsymbol{r}-\boldsymbol{r}_s|^{-1}d\tilde{z}$$

$$+ \frac{\hat{\boldsymbol{z}}}{4\pi}\int m_z(y_s'\partial_x - x_s'\partial_y)|\boldsymbol{r}-\boldsymbol{r}_s|^{-1}d\tilde{z}. \tag{A9}$$

Finally, noting that $m_z = m_0$, and that an infinitesimal segment of the string may be described as $d\boldsymbol{\ell} = (x_s'\hat{\boldsymbol{x}} + y_s'\hat{\boldsymbol{y}} + \hat{\boldsymbol{z}})d\tilde{z}$, one may further simplify Eq.(A9) to arrive at

$$\boldsymbol{A}_S(\boldsymbol{r}) = \frac{m_0}{4\pi}\int_{\infty}^{z_0}(\boldsymbol{\nabla}_r|\boldsymbol{r}-\boldsymbol{r}_s|^{-1}) \times d\boldsymbol{\ell}. \tag{A10}$$

The above expression of $\boldsymbol{A}_S(\boldsymbol{r})$ has the expected form of an integral over the vector potentials produced by elemental dipoles whose continuous arrangement constitutes the string. Jackson rightly uses Eq.(A10) as the starting point of his analysis in [2]. Our lengthy derivation of Eq.(A10) has not uncovered any problems with this straightforward integration of the contributions by infinitesimal dipoles to the vector potential at the observation point $\boldsymbol{r}$. The exercise is nevertheless worthwhile considering that it is not *a priori* obvious that the arbitrary turns and twists of a tightly-wound solenoid will *not* cause a leakage of its internal magnetic flux. Our rigorous treatment of the long thin string thus confirms that it is possible to avoid producing magnetic charges along the string without constraining its geometric configuration.

# Appendix B

Use $\nabla \times (\psi A) = \nabla\psi \times A + \psi \nabla \times A$ in going from Eq.(6a) to Eq.(6b) and also from Eq.(6e) to Eq.(6f).

Use $\oint_c \psi d\ell = -\int_s \nabla\psi \times d\boldsymbol{\sigma}$ in going from Eq.(6c) to Eq.(6d).

Use $\nabla \times (\nabla \times A) = \nabla(\nabla \cdot A) - \nabla^2 A$ in going from Eq.(6f) to Eq.(6g).

Use $\nabla \cdot (\psi A) = A \cdot \nabla\psi + \psi \nabla \cdot A$ in going from Eq.(6g) to Eq.(6h).

Use $\nabla^2 |r^{-1}| = -4\pi \delta_3(\boldsymbol{r})$ in going from Eq.(6h) to Eq.(6i).

In Eq.(6j), the solid angle subtended by the surface element $d\boldsymbol{\sigma}$ located at $\boldsymbol{r}_s$ and viewed from $\boldsymbol{r}$ is $d\Omega(\boldsymbol{r}; \boldsymbol{r}_s)$.

When the observation point $\boldsymbol{r}$ resides outside the surface $s$, the second integral in Eq.(6j) vanishes.

In Eq.(6k), $r_\perp$ is the perpendicular distance from the observation point $\boldsymbol{r}$ to the surface $s$. As $\boldsymbol{r}$ approaches $s$, the unit-vector $\hat{\boldsymbol{r}}_\perp$ coincides with the local surface-normal.

**Author's biography**. Masud Mansuripur (PhD, 1981, Electrical Engineering, Stanford University) is Professor and Chair of Optical Data Storage at the College of Optical Sciences of the University of Arizona in Tucson. He is the author of "*Introduction to Information Theory*" (Prentice-Hall, 1987), "*The Physical Principles of Magneto-Optical Recording*" (Cambridge University Press, 1995), "*Classical Optics and its Applications*" (Cambridge University Press, 2002, second edition 2009, Japanese translation 2006 and 2012), and "*Field, Force, Energy and Momentum in Classical Electrodynamics*" (Bentham e-books, 2011). A Fellow of OSA and SPIE, he is the author or co-author of nearly 250 technical papers in the areas of optical data recording, magneto-optics, optical materials fabrication and characterization, thin film optics, diffraction theory, macromolecular data storage, and problems associated with radiation pressure and photon momentum.